\documentclass{ws-procs975x65}
\usepackage{amsmath}
\begin{document}

\title{BULK VISCOSITY IMPACT ON THE SCENARIO OF WARM INFLATION}
\author{Jose Pedro Mimoso$^*$ and Ana Nunes$^\dag$}
\address{Departmento de F\'{\i}sica e Centro de F\'{\i}sica Te\'{o}rica e Computacional\\
Universidade de Lisboa, \\
Avenida Professor Gama Pinto 2, P-1649-003 Lisboa,\\
Portugal\\
$^*$E-mail: jpmimoso@cii.fc.ul.pt\\
$^\dag$E-mail: anunes@imc.fc.ul.pt }
\author{Diego Pav\'{o}n$^\ddag$}
\address{Departmento de F\'{\i}sica,\\
 Universidad Aut\'{o}noma de Barcelona, \\
 08193 Bellaterra (Barcelona),\\
Spain\\
$^\ddag$E-mail: diego.pavon@uab.es}

\begin{abstract}
The decay of the inflaton into radiation and particles during the
slow-roll suggests that these may interact with each other and
that the latter may also decay into subproducts before inflation
is completed. As a consequence, the fluid is no longer perfect and
a non-negligible bulk viscosity necessarily sets in. We write the
corresponding equations as an autonomous system and study the
asymptotic behavior, the conditions for the existence of scaling
solutions, and show that the late time effect of fluid dissipation
alleviates the depletion of matter and increases the duration of
inflation.
\end{abstract}

\bodymatter

\section{Introduction}
In the warm inflation scenario the inflaton decays during the
slow-roll into particles and radiation at such a rate, $\Gamma$,
that at the end of the accelerated expansion the temperature of
the cosmic fluid is high enough for the conventional hot big-bang
scenario to proceed right away \cite{warm}. Thus, the reheating
mechanism -typical of cool inflation \cite{cool}- can be dispensed
with altogether. Aside from this, warm inflation has other
significant
advantages: $(i)$ the slow-roll condition $\dot{\phi}^{2} \ll %
V(\phi)$ can be fulfilled for steeper potentials, $(ii)$ the
density perturbations generated by thermal fluctuations may be
larger than those of quantum origin \cite{taylor} , and $(iii)$ it
may provide a very useful mechanism for baryogenesis
\cite{baryogenesis}.

To simplify the dynamics of warm inflation many works treated the
particles created in the decay of the inflaton purely as
radiation, thereby ignoring the existence of particles with mass
in the decay fluid. In a recent paper \cite{jpad}, we took into
account the presence of particles with mass as part of the decay
products, and provided a hydrodynamical description of the mixture
of massless and non-massless particles by an overall fluid with
equation of state $p = (\gamma - 1) \rho$, where the adiabatic
index $\gamma$ lies in the interval $[1,2]$ (for pure radiation
$\gamma = 4/3$), and endowed with a dissipative bulk pressure,
$\Pi$. The latter originates quite naturally via: $(i)$
interparticle interactions \cite{interparticle}, and $(ii)$ the
decay of particles within the fluid \cite{decay}. With regard to
$(i)$, it should be noted that $\Pi$ arises spontaneously in
mixtures of different particles species, or of the same species
but with different energies -a typical instance in laboratory
physics is the Maxwell-Boltzmann gas \cite{harris}. One may think
of $\Pi$ as the internal ``friction" that sets in as a consequence
of the diverse cooling rates in the expanding mixture, something
to be expected in the matter fluid originated by the decay of the
inflaton. As for $(ii)$, it is well known that the decay of
particles within a fluid can be formally described by  a bulk
dissipative pressure -see, e.g., \cite{decay,winfried}$\,$. In
this connection, it has been proposed that the inflaton may first
decay into a heavy boson which subsequently decays in two light
fermions \cite{boson}.

In this short Communication we limit ourselves to briefly
summarize the salient findings of Ref. \cite{jpad}, paying special
attention to the conditions for the existence of scaling solutions
-i.e., solutions for which the energy densities of the inflaton
and the fluid (matter plus radiation) keep a constant ratio.

\section{Warm inflation with $\Pi = -3 \zeta H$ and $\Gamma =$ constant}
The autonomous system of equations can be written as,
\[
x' = x [Q -3(1+ \bar{r}\chi)]-W(\phi) y^{2}\, ,\qquad y' =
[Q+W(\phi) x]y ,
\]
\[
\chi' = \chi Q \, , \qquad \phi' = \sqrt{6}\, x\, \quad (W(\phi) =
\sqrt{3/2}\,(\partial_{\phi}V/V)\, ,\; \; \chi = \Pi/(3H^{2})\,
,\; \; \; \bar{r} = -\Gamma/(3\zeta)) \, ,
\]
\\
where $x^{2}= \dot{\phi}^{2}/(6H^{2})$, $y^{2}= V(\phi)/(3H^{2})$,
$Q = (3/2)[2 x^{2} + \gamma (1-x^{2}-y^{2})+\chi]$, and a prime
stands for derivative with respect to $\ln a$ (see Ref.
\cite{jpad} for details).

Fixed points are found at finite $\phi$, namely, $x_{\ast} =
y_{\ast} = 0, \; \phi = \phi_{\ast},\; \chi_{\ast} = -\gamma$
(which correspond to matter dominated de Sitter solutions), and
$x_{\ast} = 0, \; \gamma (1-y_{\ast}^{2}) = - \chi_{\ast}$ (de
Sitter scaling solutions), and also at $\phi \rightarrow \infty$
but these do not show scaling solutions.

\section{Warm inflation with $\Pi = -3 \zeta \rho^{\alpha} \,
H^{\beta}$ and $\Gamma = \tilde{\Gamma}(\phi) \, H^{\delta} $}

In this more general case the autonomous system reads,
\\
\[
x' = x[Q-3(1+3)]-W(\phi) y^{2}\, , \qquad y' = [Q+W(\phi)x]y \, ,
\]
\[
r' = [\sqrt{6}(\partial_{\phi}\tilde{\Gamma}/\tilde{\Gamma})x +
Q(1-\delta)]r\, ,\qquad \phi' =\sqrt{6} x\, ,
\]
\\
where $\chi = -3^{\alpha}\zeta(1-x^{2}-y^{2})^\alpha$ and
$Q=(3/2)[2x^{2} + \gamma(1-x^{2}-y^{2})+\chi]$, with $\delta >0,
\; 0 <\alpha <1$, and  $\beta >0$. The results for the scaling
solutions can found in the Table III of \cite{jpad}. The fixed
points at finite $\phi$ constitute: $(i)$ a line of fixed points
with $x = y = r = 0$ (these are matter dominated and the solutions
are unstable), and  $(ii)$ another line  $x = 0, \phi = \phi_{0},
(1-y^{2})^{1-\alpha} = 3^{\alpha}\zeta/\gamma$ for any $r \neq 0$.
In this second case, when $V(\phi)$ has a minimum, there is either
a stable node, $0 < W'(\phi_{0})<9(1+r)^{2}/(4 \sqrt{6}\,
y_{\ast}^{2})$, or a stable sink, $W'(\phi_{0}> 9(1+r)^{2}/(4
\sqrt{6}\, y_{\ast}^{2}) >0$ -here $y_{\ast}^{2} =%
1-(3\alpha\zeta/\gamma)^{1/(1-\alpha)}$. A sufficiently flat
effective potential can thus be achieved with the help of viscous
effects, whence ``assisted" inflation \cite{assisted} can be
dispensed with. As for fixed points at $\phi \rightarrow \infty$,
scaling solutions exist when $W(\phi_{\infty}) = -\sqrt{3/2}
\lambda < 0$ provided that $W(\phi) = \sqrt{6}\,(\partial_{\phi}%
\tilde{\Gamma}/\tilde{\Gamma})/(1-\delta)$.

\section{Concluding remarks}
Our main findings can be summarized as follows:
\begin{enumerate}
\item
Relevant asymptotic regimes arise in connection with maxima
and minima of $V(\phi)$, at finite $\phi$, and with the asymptotic
behavior at $\phi \rightarrow \infty$.

\item In the latter case, the existence of scaling solutions
depends on $\Gamma(\phi)$. A necessary and sufficient condition
for scaling solutions is that $\Gamma(\phi) \propto H$.

\item
 Scaling solutions exist for $\Gamma \neq 0$ regardless the
steepness of $V(\phi)$.

\item When $\Pi \neq 0$ the inflationary region takes a larger
portion of the phase space than otherwise -compare Figs. 1 and 2
of Ref. \cite{jpad}. Thus, the presence of a dissipative pressure
-which arises on very general grounds- facilitates inflation and
lends support to the warm inflationary proposal.
\end{enumerate}

The next step should be to study the scalar and tensorial
perturbations stemming from this model.

\bibliographystyle{ws-procs975x65}

\end{document}